# The critical temperature of smart meta-superconducting MgB$_2$

Shuo Tao, Yongbo Li, Guowei Chen, Xiaopeng Zhao*

Smart Materials Laboratory, Department of Applied Physics, Northwestern Polytechnical University, Xi'an 710129, People's Republic of China

*Corresponding author: Prof. Xiaopeng Zhao

Tel.: +86-29-8843-1662

Fax: +86-29-8849-1000

E-mail: xpzhao@nwpu.edu.cn

**Abstract**

Enhancing the critical temperature ($T_C$) is important not only to the practical applications but also to the theories of superconductivity. $MgB_2$ is a type II superconductor with a $T_C$ of 39 K, which is very close to the McMillan limit. Improving the $T_C$ of $MgB_2$ is challenging but significant. Inspired by the metamaterial structure, we designed a smart meta-superconductor that consists of $MgB_2$ microparticles and $Y_2O_3:Eu^{3+}$ nanorods. In the local electric field, $Y_2O_3:Eu^{3+}$ nanorods will generate electroluminescence (EL) that can excite $MgB_2$ particles, thereby improving the $T_C$ by strengthening the electron–phonon interaction. Each $MgB_2$-based superconductor doped with one of the four dopants of different EL intensities was prepared by an ex-situ process. The results showed that the addition of $Y_2O_3:Eu^{3+}$ brings about an impurity effect that decreases the $T_C$ and an EL exciting effect that increases the $T_C$. Apart from the EL intensity, the micro-morphology and degree of dispersion of the dopants also affected the $T_C$. This smart meta-superconductor provides a new method for increasing $T_C$.

**Introduction**

Since the discovery of superconductivity in Hg at approximately 4.2 K in 1991 [1]，the research on superconductivity has gone through over a hundred years. Initially, superconductors were regarded as the ideal conductor. In 1933, the discovery of the Meissner effect [2] led to the revelation that superconductors possess different magnetic properties. Although many theories have been established [3-7], the microscopic mechanism of superconductivity could not be understood until the proposal of BCS theory in 1957 [8]. On the basis of BCS theory, McMillan proposed that the $T_C$ of traditional superconductor cannot exceed 40 K [9] , which is called the McMillan limit.

The critical temperatures of superconducting materials are continuously being improved. In 1973, the $T_C$ of Nb–Ge alloy was found to be at 23.2 K [10]. This record was maintained for 13 years until 1986, when Bednorz and Müller found that the $T_C$ of the La–Ba–Cu–O system is above 30 K [11]. From then, a series of high-temperature superconductors (HTSs) were discovered continuously and the value of $T_C$ had greatly increased [12-14]. By 1993, researchers from ETH Zürich successfully developed a mercury-based superconductor whose $T_C$ was 134 K [15], which can be improved up to 164 K in a high-pressure environment [16].

In recent years, a number of new superconductors have been reported. In 2008, the Japanese Hosono's group found superconductivity in F-doped LaOFeAs [17], which opened a research boom of iron-based superconductor. In 2011, Cavalleri et al. used mid-infrared laser pulses to transform $La_{1.675}Eu_{0.2}Sr_{0.125}CuO_4$ into a transient superconductor [18]. After that, this group has investigated $La_{1.84}Sr_{0.16}CuO_4$[19], $YBa_2Cu_3O_{6.5}$ [20] and $K_3C_{60}$ [21] using the similar method and observed the behavior of superconducting transition. They believed that laser pulse led to deformed crystal structures and induced superconductivity [22]. Optical method was gradually applied to investigate superconductivity. In 2015, Drozdov et al. reported that the $T_C$ of the sulfur hydride system is as high as 203 K [23]. Although the achievable maximum $T_C$ continues to increase, most superconducting materials of practical value are low-temperature superconductors, especially type II superconductors.

In 2001, Japanese scientists discovered superconductivity in a binary compound magnesium diboride ($MgB_2$) [24]. The $T_C$ of $MgB_2$ is 39 K, which is close to the McMillan limit. Its high $T_C$, simple crystal structure, large coherence lengths, high critical current densities and fields, and transparency of grain boundaries to current promise that $MgB_2$ will be a good material for both large-scale applications and electronic devices. Improving the $T_C$ of $MgB_2$ is very important not only to its practical applications, but also to the theories of superconductivity.

$MgB_2$ is a conventional BCS superconductor. It becomes superconducting when electrons form Cooper pairs via electron–phonon interaction. B atom vibrations are more important than Mg atom vibrations for $MgB_2$ superconductivity. It is generally considered that substituting the B atom can alter the critical temperature [25,26], which is a more effective approach than substituting the Mg atom, in which case the B layers would be deformed [27]. However, partially substituting Mg with Al or B with C decreases the $T_C$ of $MgB_2$ [28-30]. The superconductivity of $MgB_2$ has been carefully studied [31-33]. Another proposed method for improving the $T_C$ is by introducing more holes to increase the carrier concentration. In 2014, Adu et al. presented for the first time an enhancement (from 33.0 K to 37.8 K) of the superconducting temperature of "dirty" $MgB_2$ by doping the single-wall carbon nanotubes [34]. Improving the superconducting critical temperature of $MgB_2$ is significant but challenging.

Metamaterial is a type of composite material with the artificial structure. The properties of metamaterials are not primarily determined by the matrix material but by artificial structures such as arrangement, scale and so on [35-37]. This special structure allows certain properties to incur a great change. Based on the biological cell structure, Qiao et al. proposed an intelligent fluid dielectric particle structure model. These intelligent particles contain multiple rare-earth doped titania particles embedded into the amorphous carbon matrix, which possess hierarchical porous structure. The efficiency of mechanical/electrical energy transformation for electrorheological fluids based on these intelligent particles was considerably improved by up to eight times [38]. Cargnello et al. studied that the conductivity of lead selenide films can be manipulated by over at least six orders of magnitude with the addition of gold nanocrystals [39].

Our group has been attempting to improve the superconducting transition temperature by chemical doping for a long time. In 2007, our group studied the effects of ZnO doping on the superconductivity and crystal structure of the (Bi, Pb)-2233 superconductor. Uniformly distributed ZnO nano-defects were introduced into the BSCCO superconductors using the nano-composite method. The performance of superconductors and distribution of ZnO in superconductors according to different doping methods were examined. The results of the standard four-probe method indicated that all the samples fabricated by different doping methods had obvious performance belonging to HTS. From the HRSEM images, the nano-ZnO (about 100 nm) defects were observed to be linearly distributed on the surface of the samples fabricated by the nano-composite doping method [40]. In the same year, our group proposed that combining inorganic EL materials with metamaterials can induce a substantial change in superconducting materials, left-handed materials, photonic crystals and so on. For example, the inorganic EL-material quantum dots used as magnetic flux pinning centers, orderly arrayed in the superconducting materials, may greatly improve the $T_C$ even to room temperature [41].

With this metamaterial structure as inspiration, $MgB_2$ superconductor was doped with electroluminescent $Y_2O_3$:$Eu^{3+}$ by an in-situ process, to synthesize a superconducting metamaterial [42]. The results showed that EL contributes to the improvement of $T_C$. The temperature dependence of the resistivity of the superconductor indicates that the $T_C$ of samples decreases when increasing the amount of doped $Y_2O_3$ nanorods, due to impurity ($Y_2O_3$, MgO, and $YB_4$). However, the $T_C$ of the samples increase with increasing amount of doped $Y_2O_3$:$Eu^{3+}$ nanorods, which are opposite to doped $Y_2O_3$ nanorods. Moreover, the $T_C$ of the sample doped with 8 wt.% $Y_2O_3$:$Eu^{3+}$ nanorods was also higher than that of the doped and pure $MgB_2$. The $T_C$ of the sample doped with 8 wt.% $Y_2O_3$:$Eu^{3+}$ nanorods is 1.15 K higher than that of the sample doped with 8 wt.% $Y_2O_3$ and 0.4 K higher than that of pure $MgB_2$ [43]. However, the dopant can react with boron to form $YB_4$ during the in-situ process. The amount of residual $Y_2O_3$:$Eu^{3+}$ is small, thus leading to a great reduction of EL function.

To avoid such reaction and further research the influence of EL on $T_C$, a smart meta-superconductor with a sandwich structure, consisting of $MgB_2$ microparticles and $Y_2O_3$:$Eu^{3+}$ nanorods, was designed in the present study. The nanorods dispersed uniformly around the microparticles. The $Y_2O_3$:$Eu^{3+}$ nanorods will generate EL that improves the $T_C$ under a local electric field. Four types of dopant ($Y_2O_3$, $Y_2O_3$:$Eu^{3+}$I, $Y_2O_3$:$Eu^{3+}$II, and $Y_2O_3$:$Eu^{3+}$III) with different EL intensity were prepared by a hydrothermal method [44]. Each $MgB_2$-based superconductor doped with one of the four dopants was prepared by an ex-situ process. The obtained XRD spectra show that the dopants do not react with $MgB_2$. The $T_C$ measurements indicated that the addition of dopant brings about an impurity effect that decreases the $T_C$ and an EL exciting effect that increases the $T_C$. Apart from the EL intensity, the micro-morphology and degree of dispersion of the dopants also affect the $T_C$.

## Model

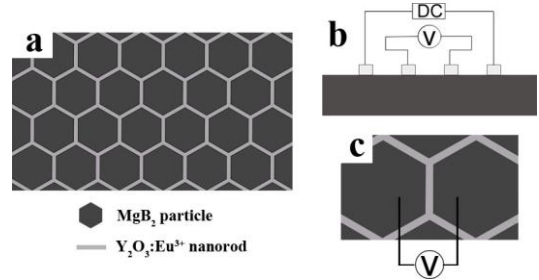

**Fig. 1** (a) The model for the smart meta-superconductor, (b) the diagram of the four-probe method, and (c) the diagram of generating local electric field.

Fig. 1a is the microstructure model of the smart meta-superconductor. The hexagons represent $MgB_2$ microparticles, and the rectangles represent the $Y_2O_3$:$Eu^{3+}$ nanorods. The nanorods disperse uniformly around the microparticles. Fig. 1b is a diagram of the four-probe method used to measure the resistance–temperature curve. Measurement of resistance was carried out in a very low temperature range in a liquid helium cryogenic system made by the Advanced Research Systems Company. The two electrodes at the two ends are connected to a direct-current (DC) source. Two intermediate electrodes are connected to a voltmeter. The DC source provides a constant current that can generate a local electric field, as shown in the Fig. 1c. A voltage drop between the two $MgB_2$ particles occurs during the measurement, in which case the two

MgB$_2$ particles act as the electrodes of the voltage source. The voltage drop produces a local electric field that can excite the intermediate nanorod to generate EL, which strengthens the electron–phonon interaction, thus improving the $T_C$. In this study, we have investigated the combined effects of the dopant and EL on the $T_C$ of MgB$_2$-based superconductors.

## Experiment

### 2.1 Preparation of MgB$_2$ power

Traditional solid-phase sintering was used to synthesize the MgB$_2$ bulk sample. Briefly, magnesium powder and boron powder were mixed at an atomic ratio of 1.1:2. The mixture was then ground for 20 min in an agate mortar in a glove box, after which the mixture was transferred into a mold and pressed into cylindrical tablets at 20 MPa for 10 min. Finally, the tablets were placed in tantalum vessels and sintered at 800 °C in a closed high-purity Ar atmosphere for 2 h. The heating rate was set to 10 °C·min$^{-1}$, and the cooling rate was set to 5 °C·min$^{-1}$. The MgB$_2$ powder was obtained by finely grinding the tablets.

### 2.2 Preparation of doped MgB$_2$-based superconductors

Four types of dopants (Y$_2$O$_3$, Y$_2$O$_3$:Eu$^{3+}$I, Y$_2$O$_3$:Eu$^{3+}$II, and Y$_2$O$_3$:Eu$^{3+}$III) of different EL intensities were prepared by a hydrothermal method, and were marked as A, B, C, and D. The molar ratio of yttrium and europium in B, C, and D was 0.95:0.05. The samples were prepared by an ex-situ process.

First, a certain quality of MgB$_2$ powder and one of the four dopants were mixed in 10 ml of ethanol to form a suspension. The suspension was transferred into a culture dish after sonicating for 30 min. Then, the dish was placed in a vacuum oven for 1 h at 60 °C. The resultant black powder was pressed into tablets. Finally, the tablets were placed in tantalum vessels and annealed at 800 °C for 1 h. The heating rate was set to 10 °C·min$^{-1}$, and the cooling rate was set to 5 °C·min$^{-1}$. Afterwards, the final products were obtained. For convenience of description, symbols were used to represent the samples. For example, the symbol A1 denotes a MgB$_2$-based superconductor in which dopant A accounts for about 1 wt.%, and so on. Furthermore, a pure MgB$_2$ sample was likewise produced by the ex-situ process, which was marked as P [45].

## Results and discussion

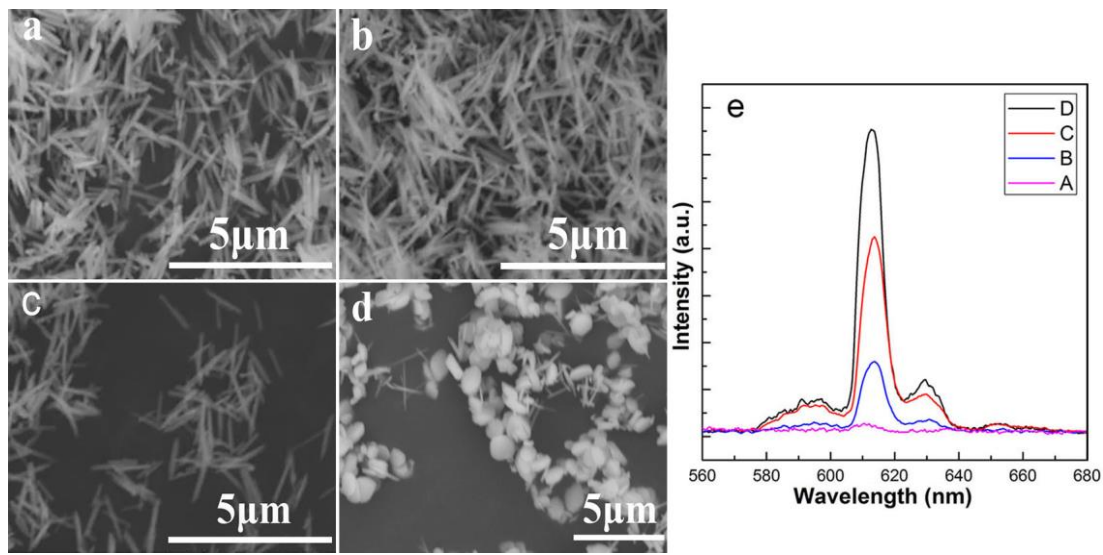

**Fig. 2** SEM images of (a) $Y_2O_3$, (b) $Y_2O_3$:$Eu^{3+}$I, (c) $Y_2O_3$:$Eu^{3+}$II, and (d) $Y_2O_3$:$Eu^{3+}$III; and (e) their corresponding electroluminescence curves.

Figs. 2a–d are the SEM images of dopants A, B, C, and D. $Y_2O_3$ nanorods of about 2 μm length and 150 nm diameter are shown in Figs. 2a–c. The nanorods in these three pictures are very similar. Fig. 2d is the SEM image of dopant D. Apart from some nanorods, many micron-sized blocks were observed. The addition of europium did not change the micro-morphology of the dopants; however, it changed their EL intensities, as shown in the EL spectrum of the four dopants in Fig. 2e. $Y_2O_3$ is not an electroluminescent material in itself. Europium act as the luminescence centre resulting in $Y_2O_3$:$Eu^{3+}$ a strong phosphor. The EL intensity of dopant C is nearly three times as high as that of dopant B. The EL intensity of D is the highest in this spectrum. All the EL curves show that the strongest peak was centered at 613 nm, with a half-width of about 8 nm, which suggests superior monochromaticity.

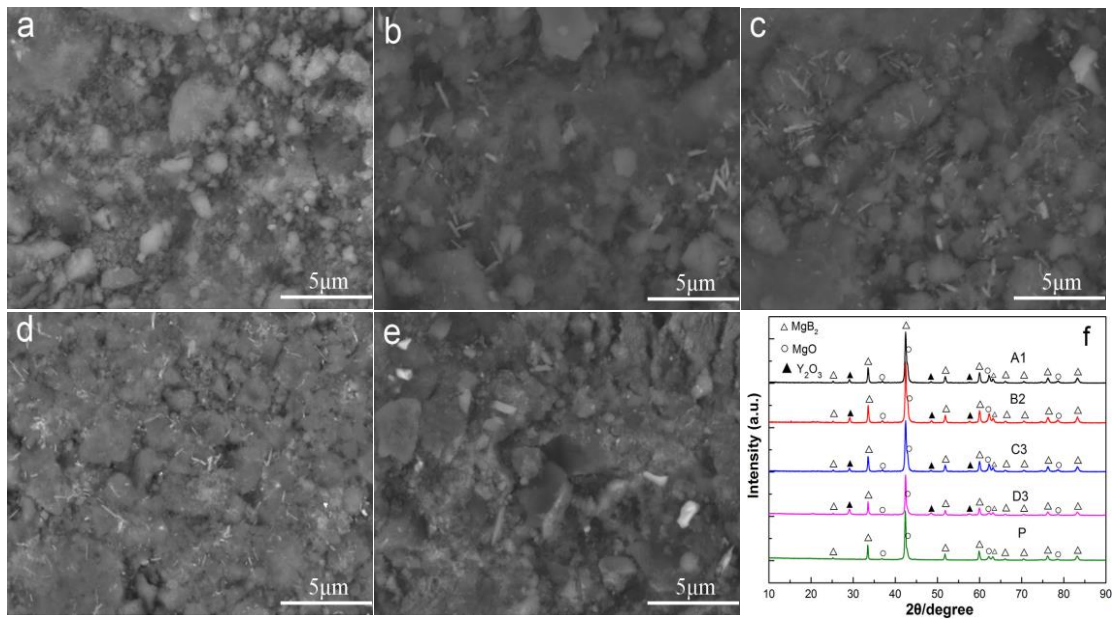

**Fig. 3** SEM images of (a) P, (b) A1, (c) B2, (d) C3, and (e) D3; and (f) the corresponding XRD spectrum.

Fig. 3 shows the SEM images (a–e) and XRD spectrum (f) of the partial samples. The SEM images show that the $MgB_2$ particles are mainly about 1–3 μm and irregularly shaped. $Y_2O_3$ nanorods are also observed (b–e). Besides some nanorods, many $Y_2O_3$ blocks are present, as shown in Fig. 3e. The dopants do not uniformly disperse in $MgB_2$. The four dopants are believed to not react with $MgB_2$ during this ex-situ process. Fig. 3f is the corresponding XRD spectrum. The XRD results show that the main phase of the samples is $MgB_2$. We can see the existence of $Y_2O_3$ phase in the doping samples. Combined with the SEM results, the dopants are confirmed to be unreactive to $MgB_2$. The XRD patterns also indicate that the samples doped with $Y_2O_3:Eu^{3+}$ are very similar to that doped with $Y_2O_3$ for the effective incorporation of Eu into the lattice of $Y_2O_3$. In addition, MgO was present in all the samples, which is consistent with other reports. The SEM images and the XRD spectra of the other samples with different dopants of different amounts are similar.

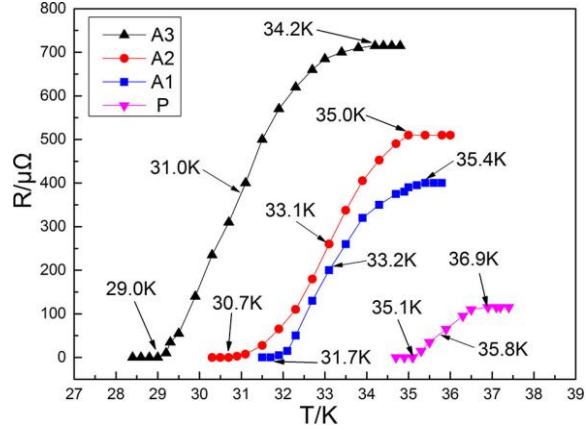

**Fig. 4** Temperature-dependent resistivity of pure MgB$_2$ (▼), and MgB$_2$ doped with 1 wt.% Y$_2$O$_3$ (■), 2 wt.% Y$_2$O$_3$ (●) and 3 wt.% Y$_2$O$_3$ (▲).

Fig. 4 presents the temperature dependence of the resistivity (R–T) of the pure MgB$_2$ and MgB$_2$ doped with Y$_2$O$_3$. The purple curve shows the R–T curve of pure MgB$_2$ (P). The $T_C$ of pure MgB$_2$ is 35.8 K. The onset ($T_C^{on}$) and offset ($T_C^{off}$) critical temperatures of pure MgB$_2$ are 36.9 and 35.1 K, which are lower than the critical temperature of pure MgB$_2$ made by the in-situ process. The main reason is that the samples were exposed to air longer in the ex-situ process compared with the in-situ process, resulting in the increment in MgO content that, in turn, decreased the critical temperature. The $T_C$ of the doping samples are lower than that of pure MgB$_2$ and decreases with increasing Y$_2$O$_3$ content. This result means that the addition of Y$_2$O$_3$ brings about an impurity effect that can decrease the critical temperature.

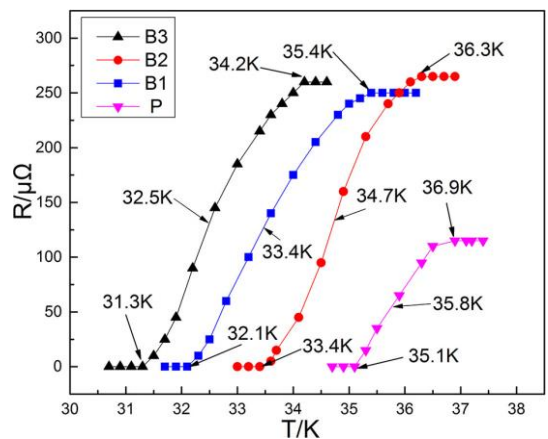

**Fig. 5** Temperature-dependent resistivity of pure MgB$_2$ (▼), and MgB$_2$ doped with 1 wt.% Y$_2$O$_3$:Eu$^{3+}$I (■), 2 wt.% Y$_2$O$_3$:Eu$^{3+}$I (●) and 3 wt.% Y$_2$O$_3$:Eu$^{3+}$I (▲).

Fig. 5 presents the temperature dependence of the resistivity of pure MgB$_2$ and MgB$_2$ doped with Y$_2$O$_3$:Eu$^{3+}$I. The $T_C$ of pure MgB$_2$ is the highest because of that the impurity effect decreases the $T_C$. However, the $T_C$ does not always decrease with increasing dopant B content. The critical temperature of B2 is higher than that of B1, which is not consistent with the impurity effect. Dopant B is thus suggested to bring another effect, apart from that of impurity effect, which contributes to the improvement of the critical temperature.

Compared with dopant A, there is little europium in dopant B. Europium, which hardly changes the morphology of Y$_2$O$_3$ makes dopant B a strong EL system (Fig. 2). We believe that EL is the key to improving the critical temperature. As shown in Fig. 1c, the current used to measure the R–T curve produces the local electric field that excites the EL of dopant B. In turn, EL excites the crystal lattice vibration, strengthening the electron–phonon interaction and resulting in the improvement of critical temperature. We call this phenomenon as an EL exciting effect. There is an obvious competitive relationship between the impurity effect and the EL exciting effect. The $T_C$ of the doping sample increases first, and then decreases along with the increment of dopant B content. The EL exciting effect is stronger than the impurity effect when the content of dopant B is 2 wt.%. However, when the content increases up to 3 wt.%, the $T_C$ decreases. This behavior is a result of the nanorods being unable to disperse uniformly in the sample. The increment of content hardly enhances the EL exciting effect but greatly increases the impurity effect.

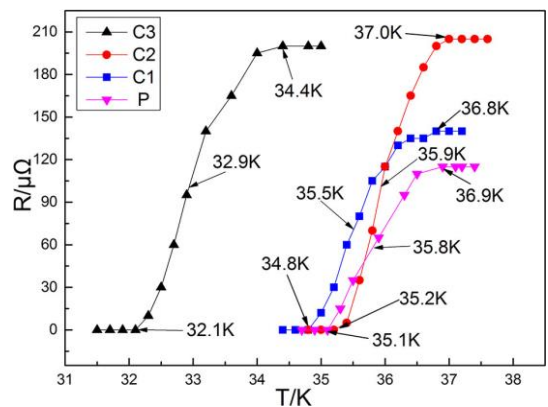

**Fig. 6** Temperature-dependent resistivity of pure MgB$_2$ ( ▼ ), and MgB$_2$ doped with 1 wt.% Y$_2$O$_3$:Eu$^{3+}$II ( ■ ), 2 wt.% Y$_2$O$_3$:Eu$^{3+}$II ( ● ) and 3 wt.% Y$_2$O$_3$:Eu$^{3+}$II ( ▲ ).

To further study the impurity effect and EL exciting effect, we produced samples doped with Y$_2$O$_3$:Eu$^{3+}$II. The critical temperatures of these samples are shown in Fig. 6. As seen from the results, doping does not always decrease the critical temperature.

The $T_C$ of C2 is 35.9 K, which is even higher than the $T_C$ of the pure MgB$_2$. The increment is not obvious, but it is of great significance. Moreover, compared with the $T_C$ of A2 in which no EL exciting effect occurs, the $T_C$ of C2 has been increased greatly by 2.8 K. This result suggests that the introduction of the EL exciting effect may be an effective method for improving the critical temperature. We also notice that the $T_C$ of C1 (C2, C3) is higher than that of B1 (B2, B3), although they have the same Eu content. We think that the high EL intensity of dopant C, which is nearly three times as high as that of dopant B, leads to the enhancement of $T_C$. Improving EL intensity can enhance the EL exciting effect, consequently improving the critical temperature. This result also indicates that the change in $T_C$ is not induced by the addition of the rare-earth element Eu.

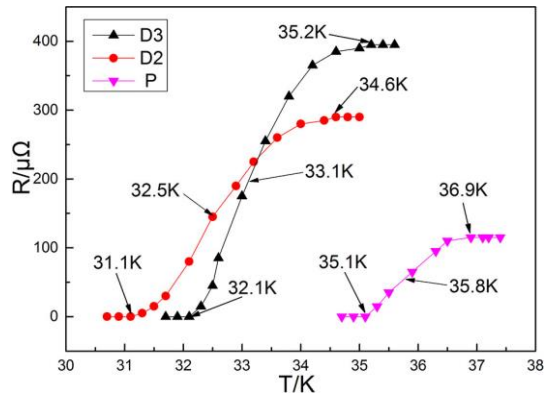

**Fig. 7** Temperature-dependent resistivity of pure MgB$_2$ (▼), MgB$_2$ doped with 2 wt.% Y$_2$O$_3$:Eu$^{3+}$III ( ● ), and 3 wt.% Y$_2$O$_3$:Eu$^{3+}$III (▲).

The critical temperatures of MgB$_2$ doped with dopant D are shown in Fig. 7. Although dopant D exhibits the strongest EL intensity among the four dopants, the critical temperatures of the corresponding samples are not higher than those of the samples doped with other dopants. This outcome is probably due to the micro-morphology of dopant D. The diameter and thickness of the Y$_2$O$_3$ block are about 2 μm and 500 nm respectively (Fig. 2d). The quantity of the Y$_2$O$_3$ blocks is much less than that of Y$_2$O$_3$ nanorods in other dopants of the same quality. The points that can produce the EL exciting effect greatly decrease, leading to the increment in the impurity effect and the reduction of the EL exciting effect. Dopants with different morphologies have different influences on the critical temperature. It also can be inferred from the result that the $T_C$ of D3 is higher than that of D2, which is different from the results shown in Figs. 5 and 6.

## Conclusion

Based on the metamaterial structure, we designed a smart meta-superconductor that consists of $MgB_2$ microparticles and $Y_2O_3$:$Eu^{3+}$ nanorods. These nanorods can generate an EL exciting effect under a local electric field produced by the current used to measure the critical temperature. The EL exciting effect contributes to the increment in $T_C$ by strengthening electron–phonon interaction. Four dopants of different EL intensities were synthesized in this study. $MgB_2$-based superconductors doped with different dopants of varying percentages were then prepared. The following conclusion can be drawn from the results. 1. The dopants can bring an impurity effect that decreases the critical temperature. The impurity effect increases rapidly with the increment in dopant content. 2. Dopants with good EL property bring about an EL exciting effect that is favorable to the improvement of the critical temperature. There is an obvious competition relationship between the impurity effect and the EL exciting effect. 3. The EL exciting effect is dependent not only on the EL intensity, but also on the doping concentration, micro-morphology of the dopants, and uniformity for dopants dispersing in the sample.

## Acknowledgements

This work was supported by the National Natural Science Foundation of China for Distinguished Young Scholar under Grant No.50025207.